\documentclass[preprint,prd,showpacs,amsmath,amssymb,amsthm,nofootinbib]{revtex4}
\usepackage[mathscr]{euscript}
\usepackage{bm}
\usepackage{textcomp}
\usepackage{graphicx}
\usepackage{subfigure}
\usepackage{overpic}
\usepackage{multirow}
\usepackage{floatrow}
\usepackage{float} 
\usepackage{amsmath,amssymb,amsthm}
\usepackage{cases}
\usepackage[colorlinks=true,linkcolor=red]{hyperref}
\newfloatcommand{capbtabbox}{table}[][\FBwidth]
\newcommand{\bea}{\begin{eqnarray}}
\newcommand{\eea}{\end{eqnarray}}
\newcommand{\beq}{\begin{equation}}
\newcommand{\eeq}{\end{equation}}

\def\/{\over}

\begin{document}

\title{Cosmological Complexity  from initial thermal state}

\author{ Jincheng Wang\footnote{J.C.Wang@hunnu.edu.cn},  Hongwei Yu\footnote{hwyu@hunnu.edu.cn} and Puxun Wu\footnote{pxwu@hunnu.edu.cn} }
\affiliation{Department of Physics and Synergetic Innovation Center for Quantum Effects and Applications, Hunan Normal University, Changsha, Hunan 410081, China 
}

\begin{abstract}
	The cosmological scalar perturbations  should satisfy the thermal distribution at the beginning of inflation since the cosmic temperature is presumably very high. In this paper,  we investigate, by the Fubini-study method, the effect of this thermal  contribution, which is characterized by a parameter $\kappa_{0}$, on the evolution of the cosmological complexity $\mathcal{C}_{FS}$ . We find that  when the thermal effect is considered,   the Universe would ``decomplex" firstly with the cosmic expansion after the mode of  the scalar perturbations exiting the horizon  in the de Sitter (dS) phase and   $\mathcal{C}_{FS}$   has a minimum about $\pi/4$. If $\mathcal{C}_{FS}$ can reach its minimum during the dS era, which requires a small $\kappa_0$ or a large e-folding number for a large $\kappa_0$, it will  bounce back to increase, and after the Universe enters the radiation dominated (RD) phase from the dS one, $\mathcal{C}_{FS}$ will decrease,  pass its minimum again, and then increase till the mode reenters the horizon. For the case of a large enough $\kappa_0$,  $\mathcal{C}_{FS}$ decreases but does not reach its minimum during the dS era, and it begins to increase after the transition from the dS phase to the RD one. When the mode reenters the horizon during the RD era, the cosmological complexity will oscillate  around about $\kappa_{0}$. These features are different from that of the initial zero-temperature case, i.e., the cosmological complexity increases during the dS phase and decreases in the RD era till the mode reenters the horizon. Our results therefore suggest that the thermal effect changes qualitatively the evolutionary behavior of the cosmological complexity. 
\end{abstract}


\maketitle
\section{Introduction}
\label{sec_in}
 According to the idea of ER=EPR~\cite{Maldacena2013},  two entangled  black holes can be regarded as  being connected by a   wormhole,  with  the growth of the wormhole volume  being described by the  entanglement entropy~\cite{Ryu2006}. 
 However, for an eternal anti-de-Sitter (AdS) black hole, 
  its entanglement entropy  saturates as it thermalizes~\cite{Hartman2013} while the size of wormwhole inside black hole is still growing,  which means  the  entanglement entropy cannot describe effectively the wormhole scale.   Complexity  has then been proposed as an extra probe to study the  characteristics of spacetime geometry. For example, Roberts, Stanford and  Susskind constructed the CV (complexity=volume) conjecture  to argue that the growth of the wormhole volume is an expected nature of quantum state complexity at the boundary~\cite{Roberts2015}. Brown et al.~\cite{Brown2016} proposed the CA (complexity=action) conjecture, which relates the quantum complexity of the holographic boundary state to the action of a Wheeler-DeWitt (WDW) patch in the AdS bulk. 

The circuit model provides a natural measure of complexity for pure states and unitaries in quantum computation~\cite{Haferkamp2022}. 
Quantum (circuit) complexity is defined as the number of unitary operators in the optimal or shortest circuit from the reference state to the target one \cite{Watrous2009}. 
 It can be obtained by employing Nielsen's geometric method~\cite{Nielsen2005,Nielsen2006,Nielsen2007,Jefferson2017} through computing the complexity between Gaussian states in the position basis with the Gaussian wave function. Later, this method was extended to form the Fubini-study metric approach~\cite{Chapman2018}, which transforms the problem of computing  complexity into finding the geodesic distance in the parametric space. Actually, the metric of the parametric space in the Fubini-study method corresponds to the covariance of the generator in the Nielsen's geometric approach. Since the properties of the Gaussian wave function can  be characterized by their exponential matrixes, the complexity can also be obtained by using the covariance matrix method~\cite{Khan2018,Hackl2018,Alves2018,Camargo2019,Ali2018,Chapman2019}.

With an upsurge of  interest in studying quantum complexity,  Bhattacharyya {\it et al.}  recently  introduced the concept of complexity into cosmology~\cite{Bhattacharyya2020}, and studied the evolution of complexity  of cosmological scalar perturbations with the cosmic expansion.  The scalar fluctuations are described by the two-mode squeezed states~\cite{Grishchuk1990}, and these states satisfy the Schr\"{o}dinger equation established from the perturbed Hamiltonian~\cite{Albrecht1993}. Through this Schr\"{o}dinger equation,  the evolution of squeezing parameters can be derived. Then,  the evolution of complexity during the very early Universe was calculated by using the Nielsen's geometric method and the covariance matrix method, respectively.

In Ref.~\cite{Bhattacharyya2020}, the initial state is chosen as an unsqueezed and zero-temperature vacuum state. This choice seems to be reasonable during inflation since the cosmic temperature $T$ satisfies $T\propto a^{-1}\propto e^{-H t}$, and thus drops rapidly to be negligible,  where $a$ is the cosmic scale factor, $t$ is the cosmic time and $H=\frac{\dot{a}}{a}$ is the Hubble parameter with an overdot denoting a derivative with respect to $t$.  However, according to the big bang theory  the cosmic temperature $T$  is very high at the beginning of inflation.   
Thus,  the initial state should be  a thermofield vacuum  state rather than the  Bunch-Davies vacuum  and the thermal effect on the cosmological complexity  needs to be considered,   and this is the  motivation of the present work.

The organization of the paper  is as follows. In Sec. \ref{sec2}, we will introduce the Hamiltonian of cosmological scalar  perturbations.  Sec. \ref{sec3} shows  the evolution of the two-mode squeezed state originating from thermal initial state. In Sec. \ref{sec4}, the cosmological complexity will be calculated by  using the Fubini-study metric approach.  We conclude at Sec. \ref{sec5}.

\section{Hamiltonian of  scalar perturbations}
\label{sec2}
We consider  the Universe filled with a perfect fluid. Thus,  the system action has the form 
\begin{eqnarray}\label{Action1}
S= \frac{1}{2} \int dt\; d^3x \sqrt{-g} [R+\mathcal{L}]\; ,
\end{eqnarray}
where $g$ is the determinant of the metric tensor, $R$  the Ricci scalar and $\mathcal{L}$ the Lagrangian density of the perfect fluid. From this action, one can derive the Einstein equation. Its $0-0$ component gives the Friedmann equation 
\begin{eqnarray}\label{FE}
			H^{2}=\frac{8\pi G_{N}}{3} \rho, 
\end{eqnarray}
where  $G_N$ is the Newton gravitational constant,  and $\rho$ is the energy density of the perfect fluid, which satisfies
\begin{eqnarray}\label{ES}
			\rho a^{3(1+w)}=\text{constant} 
\end{eqnarray}
assuming that the equation of state parameter $w$ of the perfect fluid  is a constant.   The Universe with $w=-1$ or $1/3$ corresponds to the de Sitter (dS) era or the radiation dominated (RD) era, respectively.  
Introducing the conformal time $\eta$: $a d\eta=d t$, we obtain  from Eqs.~(\ref{FE}) and (\ref{ES}) that the scale factor can be expressed as a power law  of  $\eta$
\begin{eqnarray}\label{aeta}
	a(\eta) \sim\left(\frac{\eta}{\eta_{0}}\right)^{\bar{\beta}}=\left\{\begin{array}{ll}
		-\frac{1}{H \eta} & \text {dS era} ,~~\bar{\beta}=-1  \\
		\frac{\eta}{\eta_{0}} &  \text {RD era},~~\bar{\beta}=1
	\end{array}\right.
\end{eqnarray}
in different cosmic eras, 
where $\bar{\beta}=2/(1+3w)$.

Now we consider the cosmological perturbations and are only interested in the scalar fluctuations, which are described by the curvature perturbation $\mathcal{R}$.
Then, expanding the action given in Eq.~(\ref{Action1}) into the second order, one can obtain the perturbed  action~\cite{Mukhanov1992}
\begin{align}
	S^{(2)}=\frac{1}{2} \int d\eta d^3x \left[\mu'^2+\left(\partial_i \mu\right)^2+\left(\frac{z'}{z}\right)^2\mu^2-2\frac{z'}{z}\mu' \mu\right].
\label{Eq3}
\end{align}
Here   $\mu\equiv z\mathcal{R}$ is the Mukhanov variable,  $z\equiv a \sqrt{2\epsilon}$, $\epsilon=-\dot{H}/H^2$ is one of the slow-roll parameters,  and a prime denotes a derivative with respect to $\eta$.  Note that during the  dS period the Hubble parameter $H$ is a constant and  $z=a$. From Eq.~(\ref{aeta}), one has 
\begin{eqnarray}
	\frac{z'}{z}=\frac{\bar{\beta}}{\eta}\;.
\end{eqnarray}
   
Since the two-mode case is considered when studying the cosmological complexity, we extend $\mu$ from  real to complex.  Thus, the action  Eq.~(\ref{Eq3}) indicates that the Lagrangian  density $\mathcal{L}_{\vec{k},-\vec{k}}$ in the momentum space has the form
\begin{align}\label{eq7}
	\mathcal{L}_{\vec{k},-\vec{k}}={\mu'_{k}}^\dagger\mu'_{k}-\frac{z'}{z}\left(\mu'_{k}{\mu^\dagger_{k}}+{\mu'_{k}}^\dagger\mu_{k}\right)+\left[\left(\frac{z'}{z}\right)^2-k^2\right]\mu^\dagger_{k}\mu_{k}\;,
\end{align} 
 where $\mu_{k} (\eta)$ is the Fourier transform of  $\mu(\eta,\vec{x})$  with $k$ being the comoving wave number.  
From Eq.~\eqref{eq7}, one can obtain that the Hamiltonian  has the following  symmetric form  
\begin{align}
	\mathcal{H}_{\vec{k},-\vec{k}} ={p_{k}}^\dagger {p_{k}}+\frac{z'}{2z}\left (\mu_{k}p_{k}+p_{k}\mu_{k}+\mu_{k}^\dagger p_{k}^\dagger+p_{k}^\dagger\mu_{k}^\dagger\right )+k^2\mu_{k}^\dagger\mu_{k},  
\end{align}
where $p_{k}=\partial \mathcal{L}/\partial\mu'_{k}$ is the generalized momentum.  Since $\mu_{\vec{k}}$ and $p_{\vec{k}}$ can be expressed, by introducing the usual creation and annihilation operators,  as 
\begin{align}
	\mu_{k}=\frac{1}{\sqrt{2k}}\left( a^\dagger_{-\vec{k}}+a_{\vec{k}}\right),     \;\;p_{k}=i\sqrt{\frac{k}{2}}\left( a^\dagger_{\vec{k}}-a_{-\vec{k}}\right)
\end{align}
the Hamiltonian  can be rewritten to be
\begin{align}
	\mathcal{H}_{\vec{k},-\vec{k}} =k\left(a_{\vec{k}}a^\dagger_{\vec{k}}+a^\dagger_{-\vec{k}}a_{-\vec{k}}\right)-i\frac{z'}{z}\left(a_{\vec{k}}a_{-\vec{k}}-a^\dagger_{\vec{k}}a^\dagger_{-\vec{k}}\right).
\label{Eq12}
\end{align}
The first term in Eq.~(\ref{Eq12}) is the Hamiltonian of a  free-particle, and the second one represents the interaction between scalar fluctuations and environment, which will lead to the particle creation and annihilation in pairs with the cosmic expansion. In the subhorizon limit, the first term dominates over the second one, and thus  the created $k$-mode oscillates rapidly as a harmonic oscillator. During this oscillation, since the physical wave number decays as $a^{-1}$ with the cosmic expansion, the energy of the mode decays as $a^{-1}$ too.  

\section{Squeezing evolution of two-mode thermal state}
\label{sec3}

At the beginning of inflation, which is described as a dS stage here,  it is reasonable to assume that the initial state of the scalar fluctuations is the thermofield vacuum state  since the cosmic temperature is very high.   
Furthermore, as we consider two-mode case: the particles are created in pairs with opposite momenta, we take the initial  state of the scalar perturbations as the two-mode thermofield vacuum state  which  can be defined by using   the usual creation and annihilation operators~\cite{Takahashi}
\begin{eqnarray}\label{TFD}
	|\psi_R\rangle =\exp\left [\kappa_{0}\left (a^\dagger _{\vec{k} }a^\dagger _{-\vec{k} }-a_{\vec{k} }a_{-\vec{k}} \right ) \right ]\left|0_{\vec{k}} ; 0_{-\vec{k}}\right\rangle .
\end{eqnarray}
Here $\left|0_{\vec{k}} ; 0_{-\vec{k}}\right\rangle$ is the two-mode vacuum state, and $\kappa_0$ is the value of $\kappa$ at the beginning of inflation. $\kappa$ is a parameter characterizing the thermal effect, which satisfies $\cosh\kappa=\frac{1}{\sqrt{1-e^{-\omega/T}}}$ and $\sinh \kappa=\frac{1}{\sqrt{e^{\omega/T}-1}}$, where $\omega$ is the angular frequency   of $k$-mode. For  the subhorizon mode, its $\omega$ decays as $\omega\propto a^{-1}$ due to cosmic redshift, which will result in $\kappa=\kappa_{0}$ since $T\propto a^{-1}$. 

A comparison of Eqs.~(\ref{TFD}) and (\ref{OS}) indicates that the initial state can be understood  as a result of the two-mode squeezing operator $\hat{S}_{k}(\kappa_{0},\pi/2)$ with  the squeezing angle being  $\pi/2$  and the squeezing strength $\kappa_{0}$ acting on the two-mode vacuum state~\cite{Weedbrook2012,Ferraro2005}. Thus, Eq.~(\ref{TFD}) can be re-expressed to be $|\psi_{R}\rangle= \hat{S}_{k}(\kappa_{0},\pi/2) \left|0_{\vec{k}} ; 0_{-\vec{k}}\right\rangle$.
With the cosmic evolution,  the two-mode state will be squeezed  and has the form~\cite{Grishchuk1990}
\begin{align}
	\left|\psi_{T}\right\rangle_{\vec{k},-\vec{k}}=\hat{S}_{k}\left(r_{k}, \phi_{k}\right) \hat{\mathcal{R}}_{k}\left(\theta_{k}\right) \hat{S}_{ k}\left(\kappa_{0}, \frac{\pi}{2}\right)\left|0_{\vec{k}} ; 0_{-\vec{k}}\right\rangle.\label{Eq31}
\end{align}
where $\hat{R}_{\vec{k} }$ is the two-mode rotation operator
\begin{align}
	\hat{\mathcal{R}}_{k}\left(\theta_{k}\right) \equiv \exp \left[-i \theta_{k}(\eta)\left(\hat{a}_{\vec{k}} \hat{a}_{\vec{k}}^{\dagger}+\hat{a}_{-\vec{k}}^{\dagger} \hat{a}_{-\vec{k}}\right)\right],
\end{align}
and $\hat{S}(r_{k},\phi_{k})$ is the two-mode squeezing operator
\begin{align}
		\hat{S}_{k}\left(r_{k}, \phi_{k}\right)\equiv\exp \left[r_{k}(\eta)\left(e^{-2 i \phi_{k}(\eta)} \hat{a}_{\vec{k}} \hat{a}_{-\vec{k}}-e^{2 i \phi_{k}(\eta)} \hat{a}_{-\vec{k}}^{\dagger} \hat{a}_{\vec{k}}^{\dagger}\right)\right].
\end{align}
Here $r_{k}$ denotes the squeezing strength and $\phi_{k}$ is the squeezing angle. 

Utilizing  $\hat{S}_{k}\left(\kappa_{0}, \frac{\pi}{2}\right)=\hat{S}_{k}\left(-\kappa_{0}, 0\right)$ and  the commutation relation between $\hat{\mathcal{R}}$ and $\hat{S}$~\cite{Schumaker1986}, we obtain
\begin{align}
	\begin{aligned}
		\hat{S}_{k}\left(r_{k}, \phi_{k}\right) \hat{\mathcal{R}}_{k}\left(\theta_{k}\right) \hat{S}_{k}\left(-\kappa_{0}, 0\right) &=\hat{S}_{k}\left(r_{k}, \phi_{k}\right) \hat{S}_{k}\left(-\kappa_{0}, \theta_{k}\right) \hat{\mathcal{R}}_{k}\left(\theta_{k}\right) \\
		&=\hat{S}_{k}\left(\bar{r}_{k}, \bar{\phi}_{k}\right) \hat{\mathcal{R}}_{k}\left(\bar{\theta}_{k}+\theta_{k}\right),
	\end{aligned}
\end{align}
which is  just the general two-mode squeezing-rotation operator. 
Here  three parameters $\bar{r}_{k}$, $\bar{\phi}_{k}$, and $\bar{\theta}_{k}$ are introduced to describe the characteristics of the new squeezed state, which are related to the original parameters through
\begin{align}
	\begin{array}{c}
		e^{i \bar{\theta}_{k}} \cosh \bar{r}_{k}=\cosh \kappa_{0} \cosh r_{k}+e^{2 i\left(\theta_{k}-\phi_{k}\right)} \sinh \kappa_{0} \sinh r_{k}\, , \\
		e^{i\left(2 \bar{\phi}_{k}-2 \phi_{k}+\bar{\theta}_{k}\right)} \sinh \bar{r}_{k}=\cosh \kappa_{0} \sinh r_{k}+e^{2 i\left(\theta_{k}-\phi_{k}\right)} \sinh \kappa_{0} \cosh r_{k}\, .
	\end{array}
\end{align}
 Under the occupation number representation,   Eq.~(\ref{Eq31}) can be re-expressed as (for more details see Appendix \ref{AppB})
\begin{align}
	\left|\psi_{T}\right\rangle_{\vec{k},-\vec{k}}=\frac{e^{-i\left(\theta_{k}+\bar{\theta}_{k}\right)}}{\cosh \bar{r}_{k}} \sum_{n=0}^{\infty}\left(-e^{2 i \bar{\phi}_{k}} \tanh \bar{r}_{k}\right)^{n}\left|n_{\vec{k}} ; n_{-\vec{k}}\right\rangle,\label{ST}
\end{align}
where $n_{\vec{k}}$ and $n_{-\vec{k}}$ denote the particle number of $k$ and $-k$ modes, respectively, which satisfy $n_{\vec{k}}=n_{-\vec{k}}=n$.  Taking $\bar{r}_k$ to $r_k$ and $\bar{\phi}_k$ to $\phi_k$, Eq.~\eqref{ST} reduces to the one of the zero temperature case~\cite{Albrecht1993} except for a different exponential term before the summation sign. 
 
The two-mode squeezed  state \eqref{ST} satisfies   the Schr\"{o}dinger equation
\begin{align}\label{eq19}
	i \frac{d}{d \eta}\left|\psi_{T}\right\rangle_{\vec{k},-\vec{k}}=\hat{\mathcal{H}}_{\vec{k},-\vec{k}}\left|\psi_{T}\right\rangle_{\vec{k},-\vec{k}}.
\end{align}
with the Hamiltonian given in \eqref{Eq12}. From Eq.~\eqref{eq19}, we derive the evolution equations  for the squeezing parameters 
\beq\label{rk5}
		\bar{r}_{k}^{\prime}=-\frac{z^{\prime}}{z} \cos \left(2 \bar{\phi}_{k}\right) ,
\eeq
\beq\label{phik5}
	\bar{\phi}_{k}^{\prime} =-k+\frac{z^{\prime}}{z} \operatorname{coth}\left(2 \bar{r}_{k}\right) \sin \left(2 \bar{\phi}_{k}\right), 
\eeq
\beq
	\left(\bar{\theta}_{k}+\theta_{k}\right)^{\prime}=k-\frac{z^{\prime}}{z} \tanh \left(\bar{r}_{k}\right) \sin \left(2 \bar{\phi}_{k}\right).
\eeq
When $\kappa_{0}=0$, one has $\bar{r}_{k}=r_{k}$, $\bar{\phi}_{k}=\phi_{k}$, and $\bar{\theta}_{k}=0$. Apparently,  $\bar{\theta}_{k}+\theta_{k}$ has no effect on the evolution of $\bar{r}_k$ and $\bar{\phi}_k$ since it only represents a global phase transformation. Thus, it will be discarded  in the following discussions. 
Since the initial state is the thermofield vacuum state, the corresponding initial values of squeezing parameters $\bar{r}$ and $\bar{\phi}$ are $\bar{r}_{k}=\kappa_{0}$ and $\bar{\phi}_{k}=\pi/2$, respectively. There are no exact solutions for  Eqs.~(\ref{rk5}) and (\ref{phik5}). So,   in the following we will start with an approximate analysis and then perform numerical calculations to study the evolution of the squeezing parameters. 

\subsubsection{Analytical  Approach}

The evolution of the squeezing parameters during the dS era will be discussed here approximately. For a co-moving wave mode which exits the horizon in  the dS period, it satisfies $k=a_*H$, where $a_{*}$ denotes the value of the scale factor at the time of the $k$-mode passing through the horizon. Equations (\ref{rk5}) and (\ref{phik5}) then can be re-expressed as
\beq
		\frac{d\bar{r}_{k}}{da}=-\frac{\cos{\left(2\bar{\phi}_{k}\right)}}{a}\, ,\label{Eq42}
\eeq
and
\beq
	\frac{d\bar{\phi}_{k} }{da} =-\frac{a_{*}}{a^2}+\frac{1}{a} \coth{\left(2 \bar{r}_{k}\right)}\sin{\left(2\bar{\phi}_{k}\right)}\, .\label{Eq43}
\eeq
If a large squeezing  strength $\bar{r}_{k}$ is considered, we have $\coth{\left(2 \bar{r}_{k}\right)}\approx 1$. Thus,  in the subhorizon limit ($a\ll a_{*}$),  Eqs.~\eqref{Eq42} and \eqref{Eq43} can be simplified to be
\begin{align}
	\begin{aligned}
		\frac{d\bar{r}_{k}}{da}&=-\frac{\cos{\left(2\bar{\phi}_{k}\right)}}{a},\\
		\frac{d\bar{\phi}_{k} }{da}&\simeq-\frac{a_{*}}{a^2} \,.
	\end{aligned}
\end{align}
 It is easy to obtain a solution of  $\bar{\phi}_{k}$
\begin{align}\label{Eq28}
	\bar{\phi}_{k}\simeq \frac{a_{*}}{a}-\frac{a_{*}}{a_{0}}+\frac{\pi}{2}.
\end{align}
Here $a_{0}$ is the initial value of the scale factor.  Since the scale factor increases exponentially during the dS period, Eq.~\eqref{Eq28} shows that  the squeezing angle $\bar{\phi}_{k}$ decreases rapidly before exiting the horizon, which  results in the high frequency oscillation of $\cos{\left(2\bar{\phi}_{k}\right)}$, and it finally reaches a stable value: $-\frac{a_*}{a_0} +\frac{\pi}{2}$. As the  integral of the cos function with the high frequency oscillation is zero,  $\bar{r}_{k}$ is almost  constant in the subhorizon limit.  

When  $a\gg a_{*}$, which corresponds to the superhorizon limit, Eqs.~(\ref{Eq42}) and (\ref{Eq43}) can be reduced  to
\bea
		\frac{d\bar{r}_{k}}{da}&=&-\frac{\cos{\left(2\bar{\phi}_{k}\right)}}{a},\\
		\frac{d\bar{\phi}_{k} }{da} &\simeq & -\frac{a_{*}}{a^2}+\frac{\sin{\left( 2\bar{\phi}_{k}\right)}}{a} 
	\eea
when $r_k$ is very large. After setting  $\sin{\left(2\bar{\phi}_{k}\right)}=2a_{*}/a$,   it is not difficult to find  a set of approximate stable  solutions 
\beq
\bar{r}_{k} \approx \bar{r}_{k_*}+\ln \frac{a}{a_{*}} \, , \label{Eq48}
\eeq
\beq
	\quad\quad\quad\bar{\phi}_{k} \approx -\frac{a_{*}}{a}+\left (\frac{1}{2}+2n\right)\pi,\; n\in Z \, .\label{Eq49}
\eeq
Here $\bar{r}_{k_*}$ is the value of squeezing strength at the time of the $k$ mode crossing the horizon. 
Equation \eqref{Eq48} shows that $\bar{r}_{k}$ increases proportionally to the logarithm of the scale factor and thus proportionally to the e-folding number: $\bar{r}_{k}\propto N_{e}$.

\subsubsection{Numerical Analysis}

\begin{figure}[htbp]
	\centering
	\includegraphics[width=0.45\textwidth]{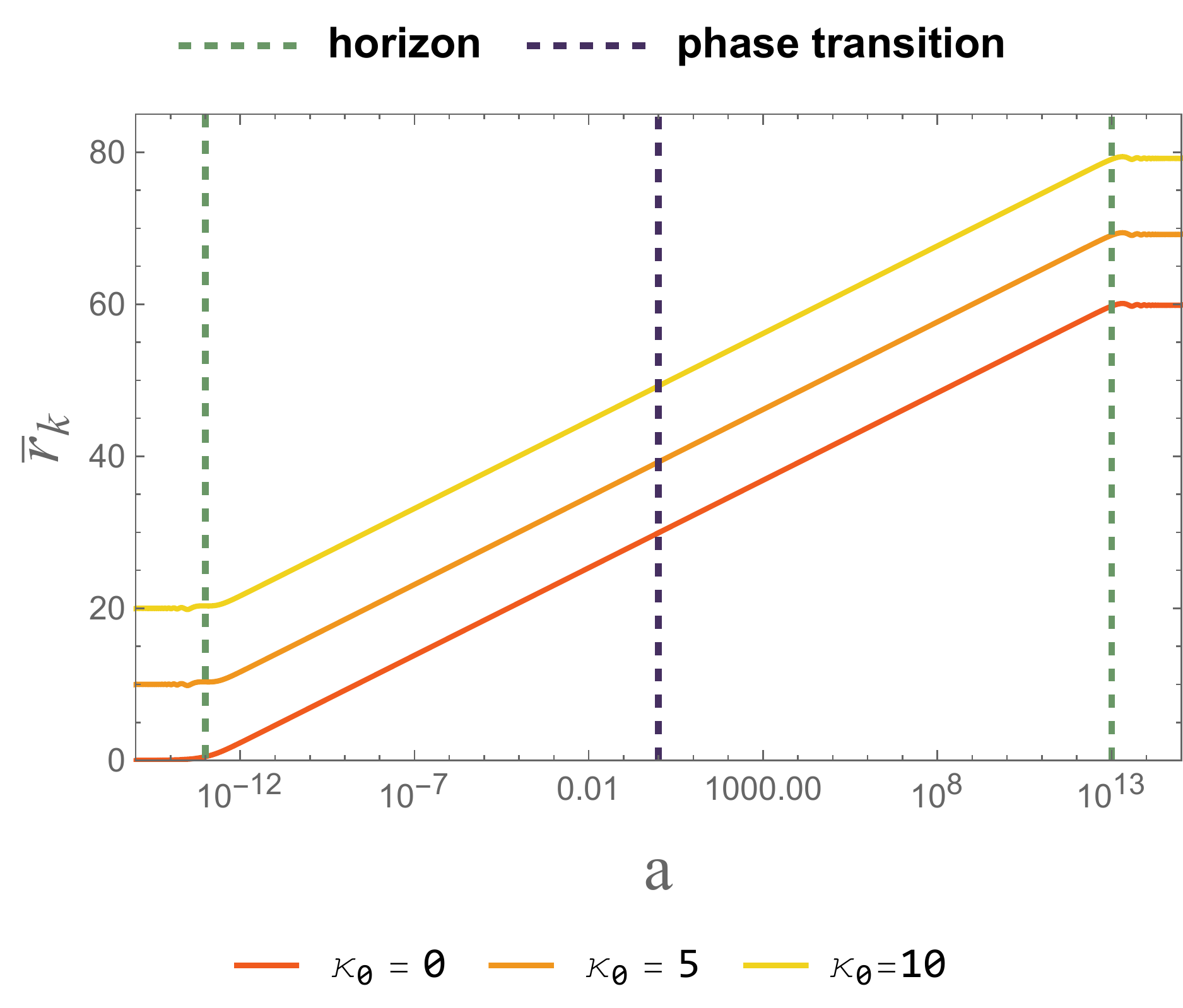}
	\caption{The evolution of the squeezing strength $\bar{r}_{k}$ as a function of the scale factor $a$ with different values of $\kappa_{0}$. The comoving mode exits the horizon at $a=10^{-13}$ in the dS period, and re-enters the horizon at $a=10^{13}$ in the RD period. The  purple dashed line represents the phase transition from the dS period to the RD one.}\label{fig1}
\end{figure}

\begin{figure}[htbp]
	\centering
	\includegraphics[width=0.45\textwidth]{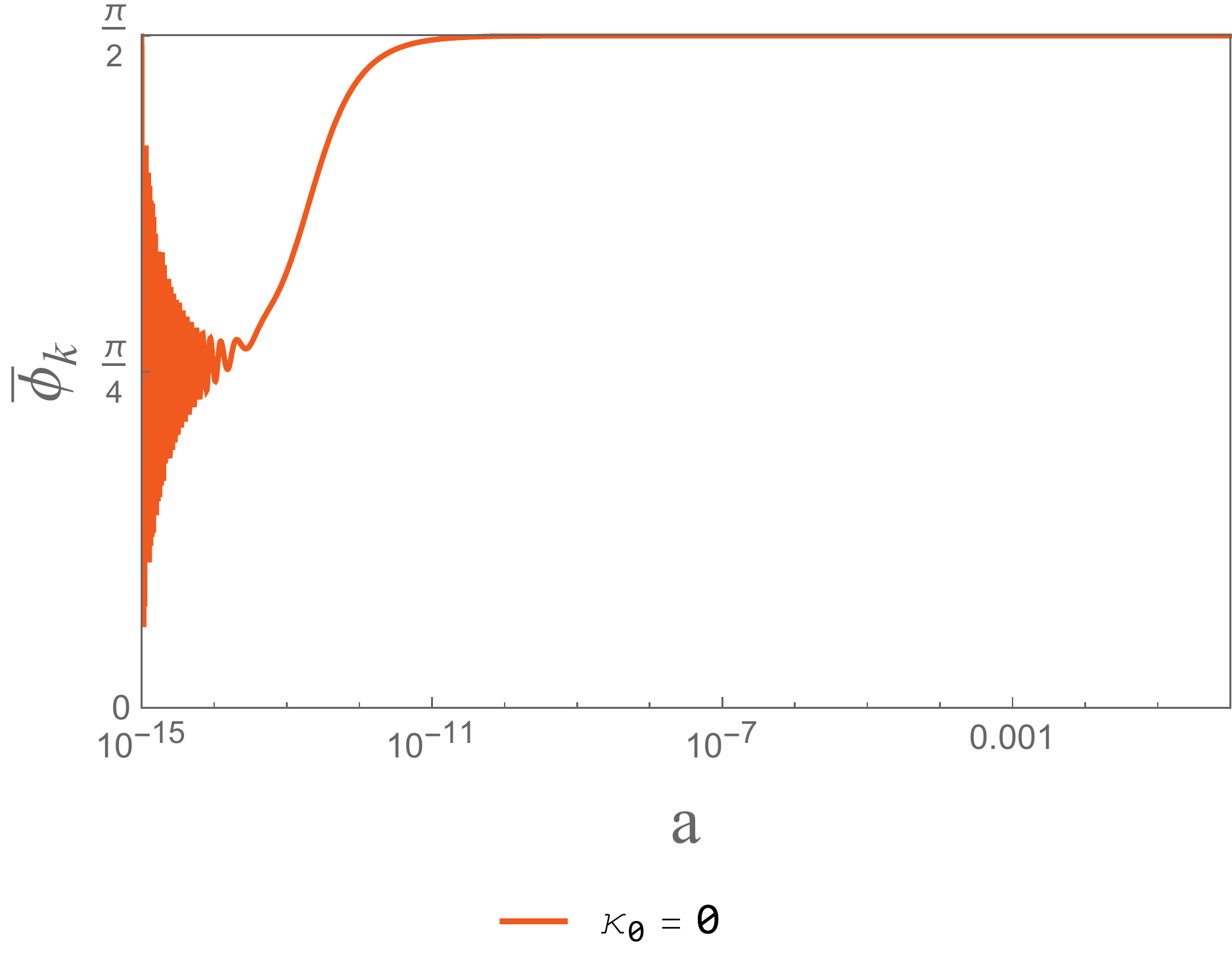}
	\includegraphics[width=0.45\textwidth]{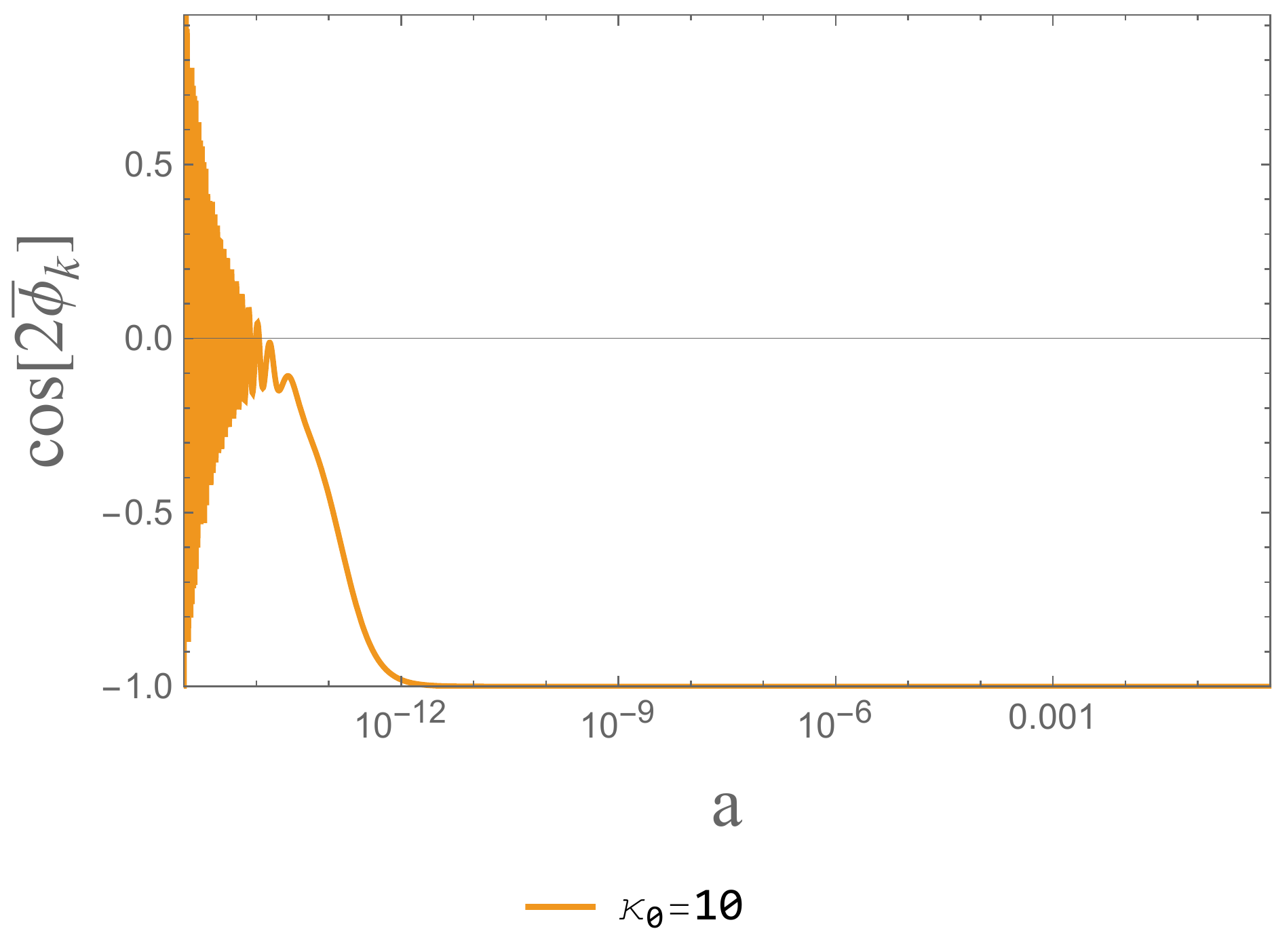}
	\caption{The evolution of the squeeing angle $\bar{\phi}_{k}$ against  $a$ during the dS era. The mode exits the horizon at $a=10^{-13}$. }\label{fig2}
\end{figure}

\begin{figure}[htbp]
	\centering
	\includegraphics[width=0.45\textwidth]{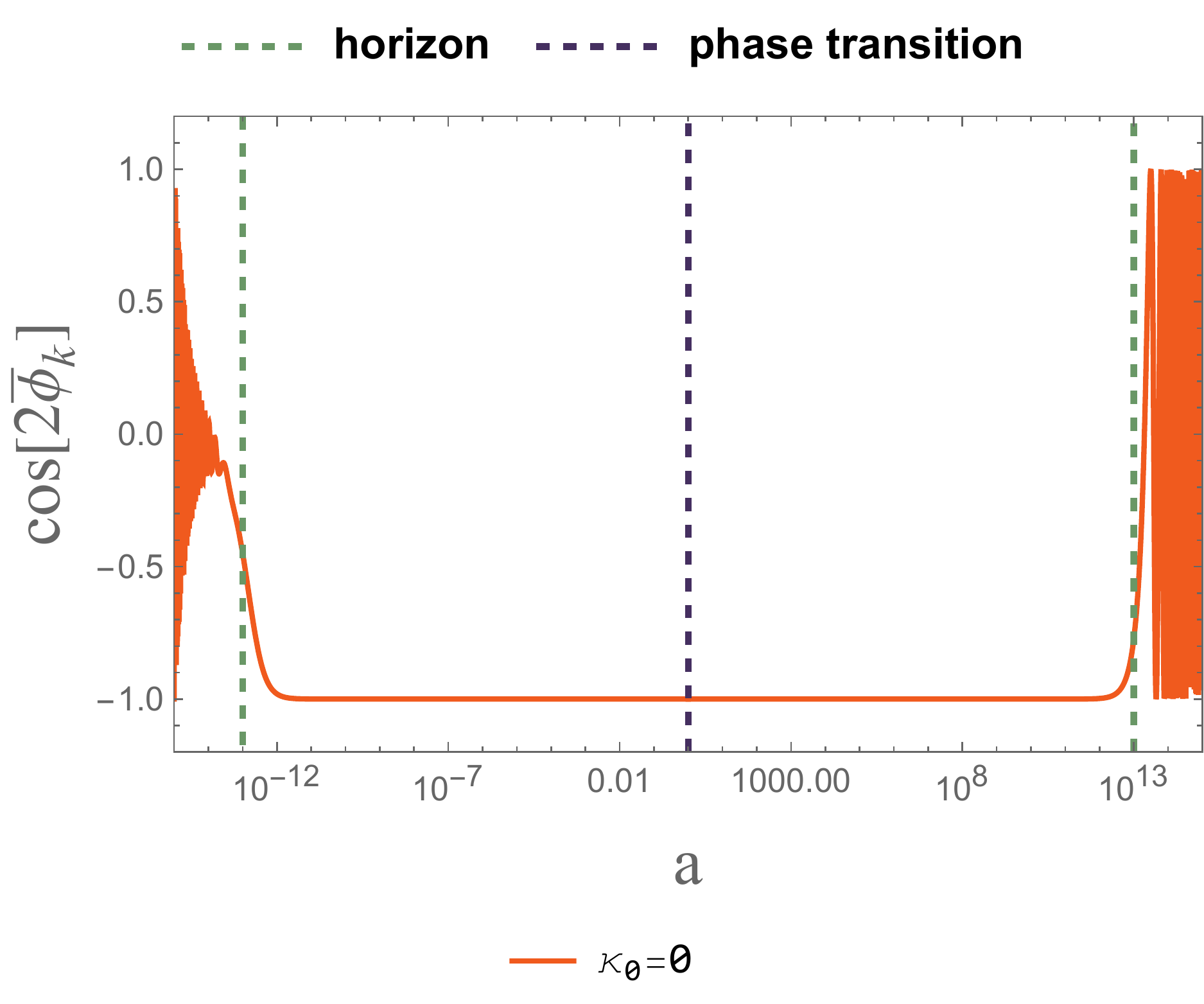}
	\includegraphics[width=0.45\textwidth]{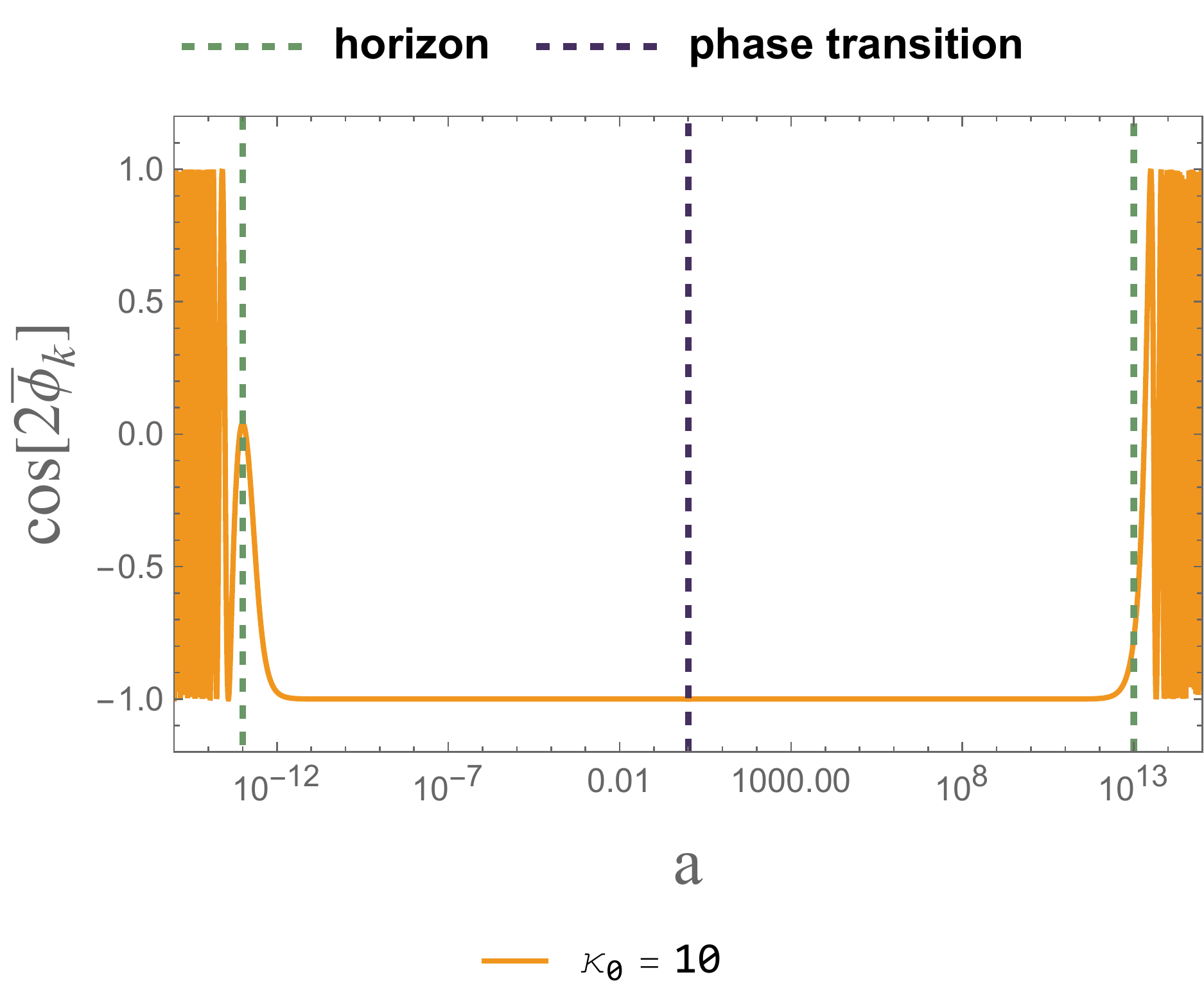}
	\caption{The evolution of $\cos(2\bar{\phi}_{k})$ from the dS phase to the RD era.    The mode exits the horizon when $a=10^{-13}$ in the dS period, and re-enters the horizon at $a=10^{13}$ in the RD period. The  purple dashed line represents the phase transition from the dS period to the RD one. }\label{fig3}
\end{figure}

To figure out the evolutionary behavior of the squeezing parameters in more detail, we must use the numerical method to solve  Eqs.~(\ref{Eq42}) and (\ref{Eq43}). The results are shown  in Figs.~(\ref{fig1}, \ref{fig2}, \ref{fig3}).  Figure (\ref{fig1}) shows the evolution of the squeezing strength $\bar{r}_{k}$ from the dS era to the RD phase with different values of $\kappa_{0}$.  Different values of $\kappa_{0}$ correspond to different initial thermofield vacuum states and $\kappa_{0}=0$ means that the initial state is the Bunch-Davies vacuum, which has been studied in~\cite{Bhattacharyya2020}.   We assume that the mode is inside the horizon initially. This figure shows clearly that  the squeezing strength is almost constant when the mode is inside the horizon, and it  increases linearly with the e-folding number $N_e$. These results are consistent with the previous analytical analysis. After re-entering the horizon, the squeezing strength will be freezed  at a constant too.  It is easy to see that  the evolutionary character of the squeezing strength is independent of the value of $\kappa_{0}$.  


Figure (\ref{fig2}) displays the evolution of the squeezing angle $\bar{\phi}_k$ against $a$ during the dS phase. 
When $\kappa_{0}=0$, $\bar{\phi}_k$ oscillates around $\pi/4$ when the mode is inside the horizon and increases to $\pi/2$ after exiting from the horizon. While, for a nonzero $\kappa_{0}$, i.e. $\kappa_{0}=10$, the squeezing angle decreases rapidly as what is obtained from the approximate analysis, and evolves finally to a stable value, which is $-\frac{a_*}{a_0}+\frac{\pi}{2}$. We also plot the evolution of $\cos (2\bar{\phi}_k)$ in Fig.~(\ref{fig3}), which indicates that   $\cos (2\bar{\phi}_k)$ oscillates rapidly when the mode is inside the horizon, and reaches rapidly to a constant $-1$ when the mode becomes  superhorizon. The difference between $\kappa_{0}=0$ and $10$ is that before the mode exiting  from the horizon during the dS phase  the oscillating amplitude of  $\cos (2\bar{\phi}_k)$ decreases for the $\kappa_{0}=0$ case, while it is almost a constant when $\kappa_{0}=10$.  

\section{Cosmological complexity}
\label{sec4}

Complexity is defined as the number of unitary operators in the optimal or shortest circuit from the reference state  to the target one, which are  respectively the two-mode thermofield vacuum state and  the two-mode squeezed state for the system considered in this paper. To calculate the complexity by  the Fubini-study metric method~\cite{Chapman2018}, the two-mode thermofield vacuum $|\psi_R\rangle$ given in Eq. (\ref{TFD}) and the two-mode squeezed states $|\psi_{T}\rangle$  (\ref{ST}) are needed to be expressed as  the Gaussian wave functions~\cite{Martin2019}. To do so, we  introduce a set of auxiliary ``position" variables $q_{\vec{k}}$ and $q_{-\vec{k}}$, which satisfy $q_{\vec{k}}=\frac{1}{\sqrt{2k}}(a^{\dagger}_{\vec{k}}+a_{\vec{k}})$ and  $q_{-\vec{k}}=\frac{1}{\sqrt{2k}}(a^{\dagger}_{-\vec{k}}+a_{-\vec{k}})$, respectively. 
We find that the Gaussian wave functions of  $\psi_R$ and $\psi_{T}$ in the ``position" space have the forms:
\begin{align}
	\begin{aligned}
		\psi_{R}\left(q_{\vec{k}},q_{-\vec{k}}\right)=\left\langle q_{\vec{k}},q_{-\vec{k}} \mid \psi_{R}\right\rangle_{\vec{k},-\vec{k}}=e^{-i\theta_{k}} \sqrt{\frac{k}{\pi}}  e^{A\left(q_{\vec{k}}^{2}+q_{-\vec{k}}^{2}\right)- B q_{\vec{k}}q_{-\vec{k}}},
	\end{aligned}\label{Gau2}
\end{align}
and
\begin{align}
	\begin{aligned}
		\psi_{T}\left(q_{\vec{k}},q_{-\vec{k}}\right)&=\left\langle q_{\vec{k}},q_{-\vec{k}} \mid \psi_{T}\right\rangle_{\vec{k},-\vec{k}}\\
		&=e^{-i\left(\theta_{k}+\bar{\theta}_{k}\right)}  \sqrt{\frac{k}{\pi}} \frac{e^{ \bar{A} \left(q_{\vec{k}}^{2}+q_{-\vec{k}}^{2}\right)-\bar{B} q_{\vec{k}}q_{-\vec{k}}}}{\cosh \bar{r}_{k} \sqrt{1-e^{-4i\bar{\phi}_{k}}\tanh ^{2}\bar{r}_{k}}}.
	\end{aligned}\label{Gau}
\end{align}
Here $\theta$ and $\theta_{k}+\bar{\theta}_{k}$ merely affect the global phases,  the coefficients $A$ and $B$ are  functions of $\kappa_{0}$ 
\begin{align}
	{A}=\frac{k}{2}\left(\frac{\tanh^{2}\kappa_{0}+1}{\tanh ^{2} \kappa_{0}-1}\right),\;\; 
	{B}=2k\left(\frac{-\tanh\kappa_{0}}{\tanh^{2}\kappa_{0}-1}\right), 
\end{align}
and $\bar{A}$ and $\bar{B}$ are functions of squeezing strength  $\bar{r}_{k}$ and squeezing angle $\bar{\phi}_{k}$,
\begin{align}
		\bar{A}=\frac{k}{2}\left(\frac{e^{-4i\bar{\phi}_{k}}\tanh^{2}\bar{r}_{k}+1}{e^{-4i\bar{\phi}_{k}}\tanh ^{2} \bar{r}_{k}-1}\right),\;\;
		 \bar{B}=2k\left(\frac{e^{-2i\bar{\phi}_{k}}\tanh\bar{r}_{k}}{e^{-4i\bar{\phi}_{k}}\tanh^{2}\bar{r}_{k}-1}\right).
\end{align}
Apparently  when $\bar{r}_{k}=\kappa_{0}$, $\bar{\phi}_{k}=\pi/2$ and $\bar{\theta}_{k}=0$, $\bar{A}$ and $\bar{B}$ will reduce to $A$ and $B$. 
Performing the coordinate transformation
\begin{align} 
	q_{\vec{k}}=\frac{1}{\sqrt{2}}\left(q_{+}+q_{-}\right),\quad q_{-\vec{k}}=\frac{1}{\sqrt{2}}\left(q_{+}-q_{-}\right), 
\end{align}
we find that Eqs.~\eqref{Gau2} and \eqref{Gau} can be written in very simple forms as 
\begin{eqnarray}
\psi_{R} =\mathcal{N} e^{-\frac{1}{2} \tilde{M}^{a b}_{R} q_{a} q_{b}}, \quad	\psi_{T}=\mathcal{N} e^{-\frac{1}{2} \tilde{M}^{a b}_{T} q_{a} q_{b}},\label{dia}
\end{eqnarray}
where $\mathcal{N}$ is the normalization constant,  $a$ and $b$ ($\in\{+,-\}$) are dummy suffix, 
and $\tilde{M}_{T}$ and $\tilde{M}_{R}$, which  characterize all the properties of  the Gaussian states, are exponential matrixes  of target and reference states, respectively, and they have the forms 
\begin{align}
	\tilde{M}_{T}=\left(\begin{array}{cc}
		-2 \bar{A}+\bar{B} & 0 \\
		0 & -2 \bar{A}- \bar{B}
	\end{array}\right) \equiv\left(\begin{array}{cc}\Omega_{+} & 0 \\
		0 & \Omega_{-}
	\end{array}\right)\,,
	\quad     \tilde{M}_{R}=\left(\begin{array}{cc}
		\omega_{+} & 0 \\
		0 & \omega_{-}
	\end{array}\right),\label{Eq54}
\end{align}
with
\begin{align}
	\omega_{+}=k\frac{1+\tanh\kappa_{0}}{1-\tanh\kappa_{0}},\quad\omega_{-}=k\frac{1-\tanh\kappa_{0}}{1+\tanh\kappa_{0}}\, .
\end{align}
It is easy to confirm that 
\begin{align}
	\frac{\Omega_{+}}{k} \cdot \frac{\Omega_{-}}{k}=1,\quad \frac{\omega_{+}}{k} \cdot \frac{\omega_{-}}{k}=1\, .\label{Eq77}
\end{align}

We use the Fubini-study metric approach to examine the cosmological  complexity. From the Appendix \ref{AppC}, one can find out that  the complexity $\mathcal{C}_{FS}$ 
\begin{eqnarray}
	\mathcal{C}_{FS}&=& \frac{1}{2}\sqrt{ \ln^2 \left |\frac{\Omega_{+}}{\omega_{+}}\right| + \left( \tan^{-1}\frac{\text{Im} [\Omega_{+}]}{\text{Re} [\Omega_{+}]}\right)^2} 
	\nonumber\\
	&=&\frac{1}{2}\sqrt{\left(\ln \left|\frac{1+e^{-2 i \bar{\phi}_{k} }\tanh \bar{r}_{k} }{1-e^{-2 i \bar{\phi}_{k}} \tanh \bar{r}_{k} }\right|-2 \kappa_{0}\right)^{2}+\left[ \tan ^{-1}\left(\sin \left(2 \bar{\phi}_{k}\right) \sinh\left( 2\bar{r}_{k}\right)\right)\right]^{2}}\, , \label{Eq83}
\end{eqnarray}
which is different from  the result obtained from  the Nielsen's geometric method  by a factor of $\sqrt{2}$ ~\cite{Bhattacharyya2020}. Thus, the Nielsen's geometric method and the Fubini-study metric one can give the same evolutionary behaviors of the cosmological complexity.

We use firstly the analytical method to investigate   the evolution of complexity during the dS phase.  As what was done in the previous section, the case of  a large squeezing strength $\bar{r}_{k}$ is considered. Thus,  we have $\tanh \left(\bar{r}_{k}\right)\approx 1$, and  $\Omega_{+}\simeq -ik \cot {\left(\bar{\phi}_{k}\right)}$ from the definition of  $\Omega_{+}$. Then equation (\ref{Eq83}) can be reduced to 
\begin{align}
	\mathcal{C}_{FS}\simeq  \frac{1}{2}\sqrt{\left(\ln \left| \tan \bar{\phi}_{k} \right|-2 \kappa_{0}\right)^2+\left[ \tan ^{-1}\left(\sin \left(2 \bar{\phi}_{k}\right) \sinh\left( 2\bar{r}_{k}\right)\right)\right]^{2}}\, .\label{Eq84}
\end{align}
For the mode inside the horizon,  Eq.~(\ref{Eq28}) shows that the  squeezing angle $\bar{\phi}_{k}$ changes  rapidly, while $\bar{r}_{k}$ remains almost	constant, which results in that the complexity would oscillate around about $\kappa_{0}$ with high frequency.

For the superhorizon case, one can obtain that ${\rm Re} [ \Omega_{+}]\simeq0$ from Eq.~\eqref{Eq49}, which leads to  $\tan^{-1}\frac{{\rm Im}[ \Omega_{+}]}{{\rm Re} [\Omega_{+}]}\simeq \frac{\pi}{2}$.  At the same time, 
we can also find that $\sin{\left(2\bar{\phi}_{k}\right)}\approx2/\tan{\left(\bar{\phi}_{k}\right)} $ and  $\tan{\left(\bar{\phi}_{k}\right)}\approx a/a_{*}$ since $\sin{\left(2\bar{\phi}_{k}\right)}\approx 2a_*/a$. As a result,  Eq.(\ref{Eq84}) can be reduced to 
\begin{align}
	\mathcal{C}_{FS}\simeq \frac{1}{2} \sqrt{ \left[\ln \left( \frac{a}{a_{*}}\right)-2 \kappa_{0}\right]^2+\frac{\pi^2}{4} }\, .\label{slope}
\end{align}
Apparently, when $\ln \frac{a}{a_*}=2\kappa_{0}$, the complexity has a minimum which is $\frac{\pi}{4}$. Thus, with the cosmic expansion during inflation, the  mode will become superhorizon. Equation~\eqref{slope} shows clearly that  the Universe will ``decomplex" initially and then the complexity  increases after it reaches the minimum. To reach this minimum requires a large e-folding number if $\kappa_{0}$ is large.

\begin{figure}[htbp]
	\centering
	\includegraphics[width=0.45\textwidth]{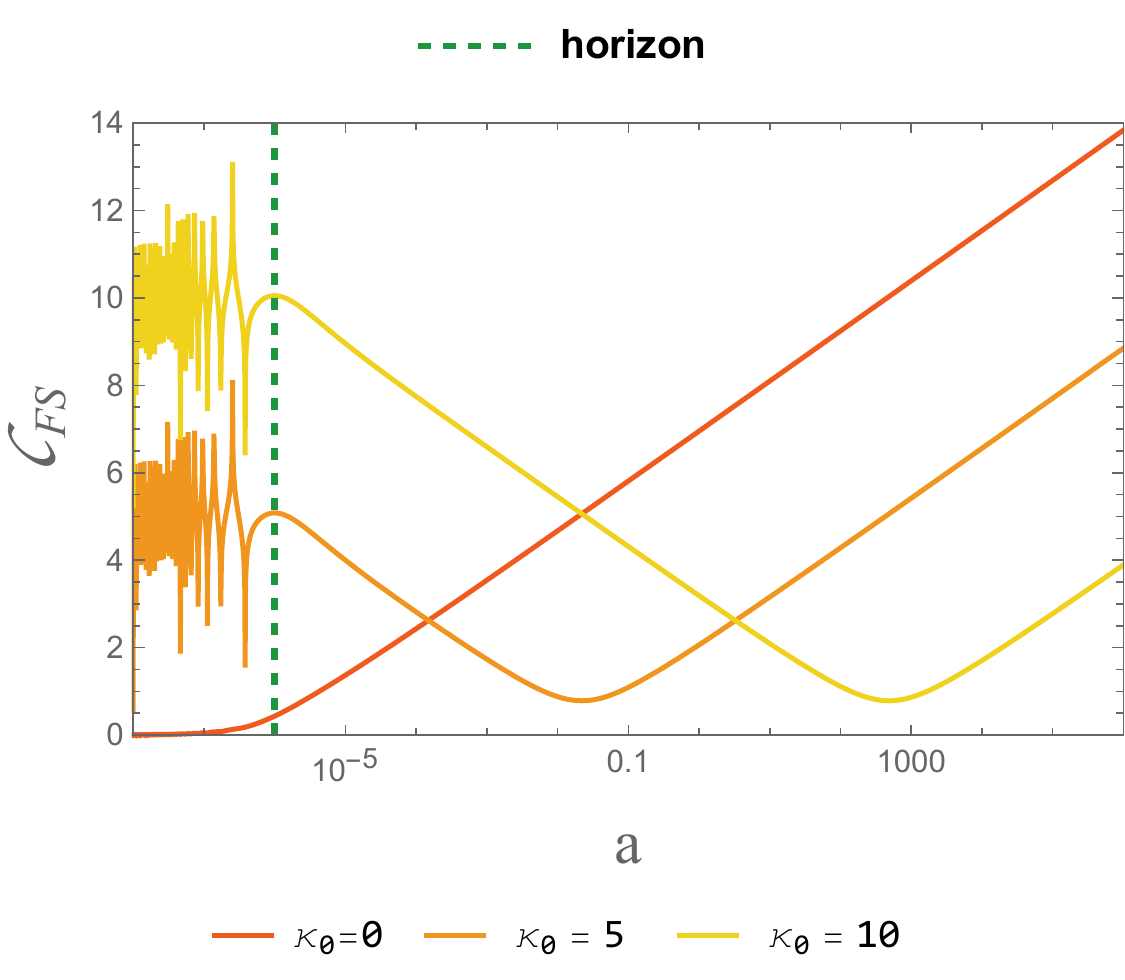}
	\caption{The evolution of  cosmological complexity $\mathcal{C}_{FS}$ against the scale factor $a$ in the dS period.  The mode exits the horizon at $a=10^{-6}$.}\label{cfs}
\end{figure}

To know in detail the evolution of $\mathcal{C}_{FS}$, we need to use numerical calculation. In Fig. (\ref{cfs}), we plot the evolutionary behaviors of  the complexity with the scale factor $a$ in the dS phase for different values of $\kappa_{0}$. The results of $\kappa_{0}=0$, which has the same trend with what were obtained in \cite{Bhattacharyya2020} from the Nielsen's geometric method, are also plotted for a comparison. It is easy to see that when the mode is inside the horizon   $\mathcal{C}_{FS}$ oscillates rapidly around about $\kappa_{0}$. It will decrease linearly with the e-folding number $N_e$ once the mode exits from the horizon. After  $\mathcal{C}_{FS}$ reaches its minimum, which is about $\pi/4$, it increases linearly with the cosmic expansion. These results are well consistent with what are obtained from analytical approach. 

\begin{figure}[htbp]
	\centering
	\includegraphics[width=0.45\textwidth]{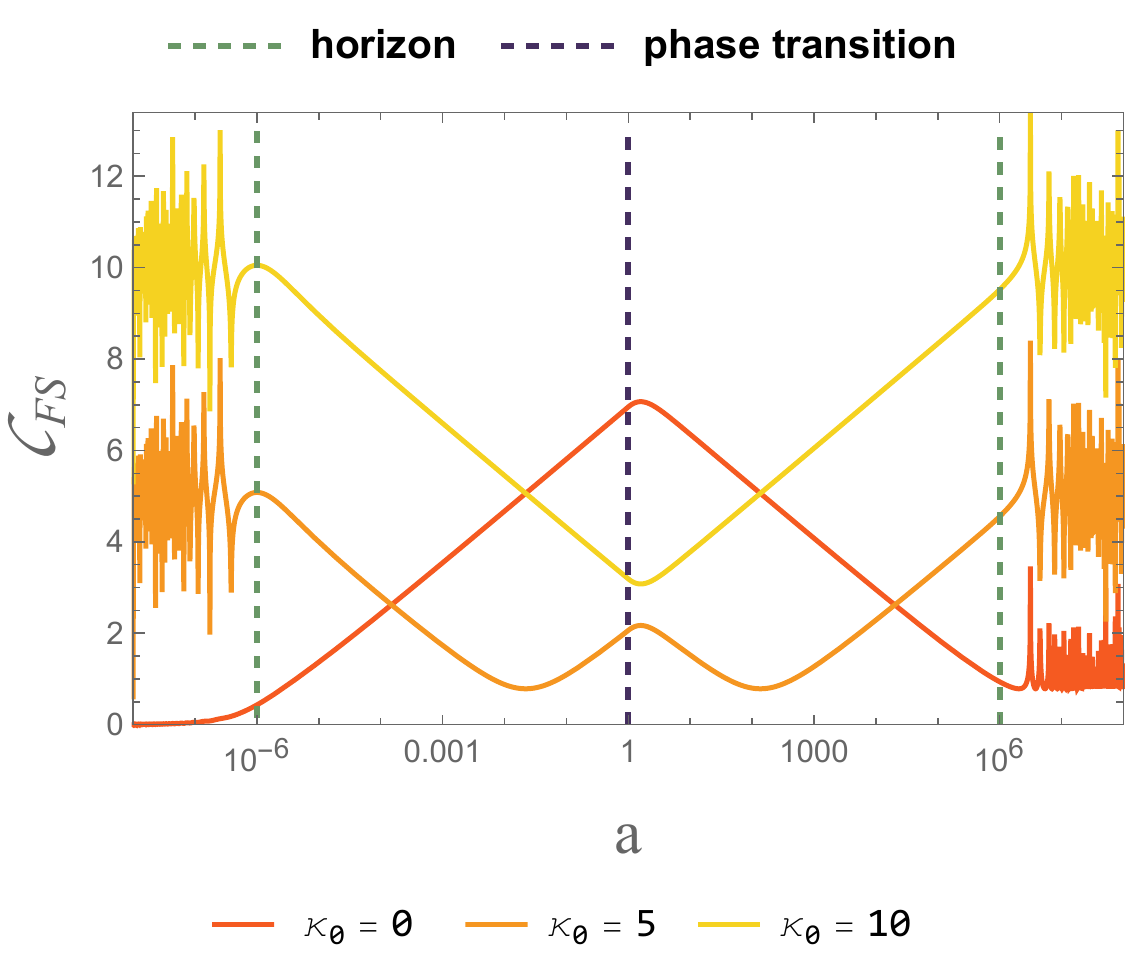}
	\caption{The evolution of  cosmological complexity $\mathcal{C}_{FS}$ against $a$ from the dS era to the RD phase. The mode exits the horizon at $a=10^{-6}$ in dS period and re-enters the horizon when $a=10^{6}$ in RD era.  The purple dashed line denotes the transition from the dS phase to the RD one.  }\label{cfsrd}
\end{figure}

Figure (\ref{cfsrd}) gives the evolutionary curves of $\mathcal{C}_{FS}$ in the dS and RD eras. During the RD era, the superhorizon mode will reenter the horizon. Apparently, the evolutionary behavior of $\mathcal{C}_{FS}$ in the case of nonzero $\kappa_{0}$ is different wholly from the one of the zero temperature case. When $\kappa_{0}=0$,  $\mathcal{C}_{FS}$  increases linearly with $N_e$ till the end of dS era after the mode crossing the horizon, and then it will decrease    when the Universe enters the RD phase.  Once the mode reenters the horizon, $\mathcal{C}_{FS}$ stops decreasing and begins to oscillate.  For a small $\kappa_{0}$ ($\kappa_{0}=5$), $\mathcal{C}_{FS}$ can reach its minimum and then bounce back to increase during the dS era. It will decrease again when the Universe enters the RD phase and then reaches its minimum.  After that, $\mathcal{C}_{FS}$ increases till the mode reenters the horizon and finally   oscillates around about $\kappa_{0}$. If $\kappa_{0}$ is large enough, i.e., $\kappa_{0}=10$, $\mathcal{C}_{FS}$ decreases with the cosmic expansion but does not reach its minimum during the dS phase.  The complexity stops decreasing at the end of dS era, and it will continue to increase during the RD era. Similar to the case of $\kappa_{0}=5$, $\mathcal{C}_{FS}$ oscillates around $\kappa_{0}$ after the mode reenters  the horizon.

\section{conclusion}
\label{sec5} 

According to the big bang theory, the Universe has a very high temperature at the beginning of inflation. Thus, the initial scalar perturbations should satisfy the thermal distribution.  We study the thermal effects on the squeezed evolution of the cosmological scalar  perturbations,  which are in the two-mode thermofield vacuum initially, and  the evolution of the cosmological complexity in the dS and RD phases. We  find that the thermal effect almost does not change  the evolutionary behavior of the squeezing strength  and the squeezing angle, but  it changes apparently the evolutionary property of the cosmological complexity $\mathcal{C}_{FS}$. In the absence of the thermal effect, which corresponds to the zero temperature case, the cosmological complexity is almost constant when the mode is inside the horizon in the dS phase. If the mode is outside the horizon, the complexity grows and decreases linearly with the e-folding number $N_e$ during the dS phase and  the RD era, respectively. However, with the thermal effect taking into account, the complexity oscillates rapidly  initially for the mode  inside the horizon and begins to decrease once the mode exits the horizon, which indicates that  the Universe would ``decomplex" firstly  with the cosmic expansion after the mode exits the horizon. If  $\mathcal{C}_{FS}$   can reach its minimum during the dS era, which requires a small $\kappa_{0}$ or a large e-folding number for a large $\kappa_{0}$, it will bounce back to increase. After the Universes enters the RD phase,   $\mathcal{C}_{FS}$ decreases, and will pass its minimum again.  For a large enough $\kappa_{0}$,  $\mathcal{C}_{FS}$ decreases but does not reach its minimum during the dS era, and it will turn to increase after the Universe enters the RD phase. This trend is  the opposite of what are obtained in the zero temperature case.  Therefore, our results suggest that the thermal effect changes qualitatively the evolutionary behavior of the cosmological complexity.

\begin{acknowledgments}
This work was supported in part by the NSFC under Grants  No. 12275080 and No. 12075084, and by the Science and Technology Innovation Plan of Hunan province under Grant No. 2017XK2019.

\end{acknowledgments}

\appendix

\section{The squeezed evolution}

In general, particles will be produced due to the change of  gravitational field  with the cosmic expansion. We assume that there are the two kinds of particle in their vacuum states initially, which are defined by the operators  $\hat{a}_{in}$ and $\hat{b}_{in}$, respectively. These vacuum states are no longer the vacuum states after the cosmic expansion. The new vacuum states can be defined by the operators $\hat{a}_{out}$ and $\hat{b}_{out}$.  These operators are related to each other through  the Bogoliubov transformations: 
\begin{align}
		\hat{a}_{in}&=e^{-i \theta} \cosh (r) \, \hat{a}_{out}+e^{-i(\theta-2 \phi)} \sinh (r) \,\hat{b}^{\dagger}_{out} , \\
		\hat{b}^{\dagger}_{in}&=-e^{i(\theta-2 \phi)} \sinh (r)\, \hat{a}_{out} +e^{i \theta} \cosh (r)\, \hat{b}^{\dagger}_{out} ,
\end{align}
which correspond to the squeezing-rotation transformation in quantum field theory
\begin{align}
	\hat{a}_{in}=\hat{R^\dagger} \hat{S^\dagger} \hat{a}_{out} \hat{S} \hat{R},\quad \hat{b^\dagger_{in}}=\hat{R^\dagger} \hat{S^\dagger} \hat{b^\dagger_{out}} \hat{S} \hat{R},
\end{align}
where 
\begin{align}
		\hat{S}(r, \phi)&=\exp \left[r \left(e^{-2 i \phi} \hat{a}_{out} \hat{b}_{out}-e^{2 i \phi} \hat{a}^{\dagger}_{out} \hat{b}^{\dagger}_{out}\right)\right],\\
		\hat{R}\left(\theta\right)&=\exp \left[-i \theta \left(\hat{a}^{\dagger}_{out} \hat{a}_{out}+\hat{b}_{out} \hat{b}^{\dagger}_{out}\right) \right].
\end{align}
The operator $\hat{S}(r, \phi)$ is  the two-mode squeezing operator with $r$ and $\phi$ denoting the squeezing strength and squeezing angle, respectively, and   $\hat{R}\left(\theta\right)$ is the two-mode rotation operator where $\theta$ represents the rotational angel of global phase. 
For an observer in vacuum state $\left|0_{in}\right\rangle$, he does not observe any number of  particles in the initial state. While in the out state the particles will be generated.  The number of $a$ particles observed in out region is 
\begin{align}
	\begin{aligned}
		\left\langle 0_{out }\left|\hat{a}^{\dagger}_{in} \hat{a}_{in}\right| 0_{out}\right\rangle &=\left\langle 0_{out }\left|\cosh ^{2} (r) \left(\hat{a}^{\dagger}_{out} \hat{a}_{out}\right)+\sinh ^{2} (r)\left(\hat{b}_{out} \hat{b}^{\dagger}_{out}\right)\right| 0_{out}\right\rangle\\&=\sinh^2 (r) .
	\end{aligned}
\end{align}
The particle creation theory can also be expressed as   that for both two observers, which can define their own vacuum, the states they are in will have a squeezing-rotational transformation relationship
\begin{align}\label{OS}
	\left|\psi_{out }\right\rangle=\hat{S}(r, \varphi) \hat{R}\left(\theta\right)\left|\psi_{in }\right\rangle.
\end{align}

\section{The occupation number representation of squeezed state}
\label{AppB}

Now, we consider a two-mode squeezed state $|\psi\rangle= \hat{S}_{k}\left(\bar{r}_{k}, \bar{\phi}_{k}\right) \hat{\mathcal{R}}_{k}\left(\bar{\theta}_{k}+\theta_{k}\right)|0_{\vec{k}};0_{-\vec{k}}\rangle$. Employing the operator transformation relation
\begin{align}
	\begin{aligned}
		\exp\left [\bar{r}_{k} \left (e^{-2i \bar{\phi}_{k}} a^\dagger _{\vec{k} }a^\dagger_{-\vec{k} }-e^{2i \bar{\phi}_{k}} a_{\vec{k} }a_{-\vec{k} } \right )\right ]
		= &\exp \left(-\tanh (\bar{r}_{k})\; e^{-2i \bar{\phi}_{k}}\, a_{\vec{k}}^{\dagger} a_{-\vec{k}}^{\dagger}\right)\\
		\cdot &\exp\left [ -\left (a^\dagger _{\vec{k} }a _{\vec{k} }+a^\dagger _{-\vec{k} }a _{-\vec{k} }+1 \right )\cdot \ln \cosh (\bar{r}_{k} )\right ]\\
		\cdot&\exp\left (-\tanh (\bar{r}_{k} )\; e^{2i \bar{\phi}_{k}}\,a _{\vec{k} }a _{-\vec{k} }\right ).
	\end{aligned}
\end{align}
one can obtain that 
\begin{align}\label{B2}
	\begin{aligned}
		\hat{\mathcal{R}}_{k}\left(\bar{\theta}_{k}+\theta_{k}\right)\left|0_{\vec{k}};0_{-\vec{k}}\right\rangle&=\exp \left[-i \left(\bar{\theta}_{k}+\theta_{k}\right)\left(\hat{a}_{\vec{k}} \hat{a}_{\vec{k}}^{\dagger}+\hat{a}_{-\vec{k}}^{\dagger} \hat{a}_{-\vec{k}}\right)\right]\left|0_{\vec{k}};0_{-\vec{k}}\right\rangle\\&=e^{-i\left(\bar{\theta}_{k}+\theta_{k}\right)}\left|0_{\vec{k}};0_{-\vec{k}}\right\rangle,
	\end{aligned}
\end{align}
and
\begin{align}\label{B3}
	\begin{aligned}
		\hat{S}_{k}\left(\bar{r}_{k}, \bar{\phi}_{k}\right)\left|0_{\vec{k}};0_{-\vec{k}}\right\rangle&=\exp\left [\bar{r}_{k} \left (e^{-2i \bar{\phi}_{k}} a^\dagger _{\vec{k} }a^\dagger_{-\vec{k} }-e^{2i \bar{\phi}_{k}} a_{\vec{k} }a_{-\vec{k} } \right )\right ]\left|0_{\vec{k}};0_{-\vec{k}}\right\rangle\\&=\frac{1}{\cosh \bar{r}_{k}} \sum_{n=0}^{\infty}\left(-e^{2 i \bar{\phi}_{k}} \tanh \bar{r}_{k}\right)^{n}\left|n_{\vec{k}} ; n_{-\vec{k}}\right\rangle.
	\end{aligned}
\end{align}
Then  Eq. (\ref{ST}) can be achieved easily by using Eqs.~\eqref{B2} and \eqref{B3}.   

\section{Calculating the quantum complexity by using the Fubini-Study Metric Approach}
\label{AppC}
The quantum fidelity can be considered as the inner product between two Gussian states~\cite{Jefferson2017,Ali2018,Chapman2018,Ruan2021},
\begin{align}\label{Fll}
	F\left(\lambda,\lambda^{\prime}\right)=\left|\left\langle\psi(\lambda)\mid\psi\left(\lambda^{\prime}\right)\right\rangle\right|\, ,
\end{align}
where $\lambda$, which is not the position basis, represents  the characteristic parameter which determines the Gaussian features. Since the Gaussian states have the orthonormal property, the quantum fidelity has a maximum value $F_{max}\left(\lambda,\lambda^{\prime}\right)=1$ at $\lambda'=\lambda$. Using this character, we take the Taylor expansion to the second-order on $F\left(\lambda,\lambda+d \lambda\right)$ at $\lambda$,
\begin{align}
	F(\lambda,\lambda+d\lambda)=1-\frac{1}{2}g_{\mu\nu}^{FS}d\lambda^{\mu}d\lambda^{\nu}+\mathcal{O}(d\lambda^3),
\end{align}
 where the second-order coefficient $g_{\mu\nu}^{FS}$ can be regarded as quantum information metric,
\beq
g_{\mu\nu}^{FS}=-\left.\frac{\partial^{2} F\left(\lambda, \lambda^{\prime}\right)}{\partial \lambda^{\mu} \partial \lambda^{v}}\right|_{\lambda^{\prime}=\lambda} =\frac{1}{2}\left(\left\langle\partial_{\mu}\psi\mid\partial_{\nu}\psi\right\rangle+\left\langle\partial_{\nu}\psi\mid\partial_{\mu}\psi\right\rangle\right)-\left\langle\partial_{\mu} \psi \mid \psi\right\rangle\left\langle\psi \mid \partial_{\nu} \psi\right\rangle.\label{Eq40}
\eeq
This metric is specially important in calculating the complexity. 
 
 Now we consider the Gaussian state on the path from which   the reference state evolves into  target state. Usually, the exponential matrix of Gaussian state can be obtained through  
\begin{align}
\tilde{M}(\sigma)=U(\sigma) \cdot \tilde{M}_{0} \cdot U^{T}(\sigma), \quad \tilde{M}_0= k I_2.\label{Eq57}
\end{align}
Here $\sigma\in [0,1]$,  $\sigma=0$ and $1$ correspond to the reference state and the target one,  respectively,   $U(\sigma)$ is the  unitary operator,  the superscript $T$ means the matrix transpose,  and $I_{2}$ is a two-dimensional identity matrix. When considering the general squeezed-rotation evolution, the unitary operator $U$ corresponds to the $G L(2, \mathbb{C})$ group~\cite{Adhikari2021}
\begin{align}
	U=e^{\xi} R(-x) S(\zeta) R(y)=e^{\xi}\left(\begin{array}{cc}
		\cos x & -\sin x \\
		\sin x & \cos x
	\end{array}\right)\left(\begin{array}{cc}
		e^{\zeta} & 0 \\
		0 & e^{-\zeta}
	\end{array}\right)\left(\begin{array}{cc}
		\cos y & \sin y \\
		-\sin y & \cos y
	\end{array}\right),\label{Eq76}
\end{align}
where $\xi,x,\zeta$ and $y$ are parameters related with the character parameters $\lambda$ of Gaussian state and they are functions of $\sigma$. Note that $\xi$ and $\zeta$ can be complex here.   
Substituting Eq. (\ref{Eq76}) into Eq. (\ref{Eq57}), we can obtain
\begin{align}
	\tilde{M}(\sigma)=k\left(\begin{array}{cc}
		e^{2\xi}\left(\cosh (2\zeta ) +\cos(2x )\sinh (2\zeta )\right) & e^{2\xi} \sin (2x ) \sinh (2\zeta ) \\
		e^{2\xi} \sin (2x ) \sinh (2\zeta ) & e^{2\xi}\left(\cosh (2\zeta ) -\cos(2x )\sinh (2\zeta )\right)
	\end{array}\right).\label{EqC6}
\end{align}
Substituting Eq.~(\ref{EqC6}) into Eq.~(\ref{dia}), one can obtain the general expression of Gaussian state, and then achieve the expression of $F(\lambda, \lambda')$ by using Eq.~(\ref{Fll}).  Thus,  from Eq. (\ref{Eq40}) we  achieve the   line element of the Fubini-Study metric, which has the form
\begin{align} \label{C7}
	ds_{F S}^{2}=|d \xi|^{2}+|d \zeta|^{2}+\sinh ^{2} (\zeta)\, d x^{2}.
\end{align}
This line element describes a parameter space, in which the reference state 	$|\psi_{R}\rangle$ and the target state $\left|\psi_{T}\right\rangle$  are just two points. The distance between these two points  is 
\begin{align}
	\mathcal{D}_{FS}=\int_{0}^{1}d\sigma\sqrt{g_{\mu\nu}^{FS}\dot{\lambda^{\mu}}\dot{\lambda^{\nu}}},
\end{align}
where $\dot{\lambda}^\mu (\sigma)=\frac{d \lambda^\mu (\sigma)}{d \sigma}$. The quantum complexity is defined as the minimum length of the geodesic between two points \begin{align}
	\mathcal{C}_{FS}\equiv\operatorname{Min}\int_{0}^{1}d\sigma\sqrt{g_{\mu \nu}^{FS}\dot{\lambda^{\mu}} \dot{\lambda^{v}}}.
\end{align}

For the system considered in this paper, the exponential matrixes ($\tilde{M}_{R}$ and $\tilde{M}_{T}$) of the reference state and the target one, which are given in  Eq. (\ref{Eq54}), satisfy $\tilde{M}_{R}=\tilde{M}(\sigma=0)$ and  $\tilde{M}_{T}=\tilde{M}(\sigma=1)$, respectively. Since all  nondiagonal elements of $\tilde{M}_{R}$ and $\tilde{M}_{T}$ are zero, we can obtain that $x=0$ or $\pi/2$ in $\tilde{M}(\sigma)$ when $\sigma=0$ and $\sigma=1$. Furthermore, both $\tilde{M}_{R}$ and $\tilde{M}_{T}$ satisfy the conditions given in Eq.~(\ref{Eq77}), which leads to that $\xi$  in $\tilde{M}(\sigma)$  must be $\xi=0$ when $\sigma=0$ and $\sigma=1$.  If choosing $x=0$ as an example, the coordinates of the reference state and the target state are $\{0, \zeta(\sigma=0),  0\}$ and  $\{0, \zeta(\sigma=1), 0\}$, respectively,  in parametric space.  For the metric given in Eq.~(\ref{C7}), the geodesic length  will be  shortest if the line is along $\xi(\sigma)=0$  and $x(\sigma)=0$, which means that the shortest  path is a straight line along $\zeta$ axis. 
Thus, the shortest distance  between the reference and target states in parameter space is
\begin{align}
	\mathcal{C}_{FS}=|\zeta(\sigma=1)-\zeta(\sigma=0)|.
\end{align}
Since $\zeta(\sigma=1)$ may be a complex number while   $\zeta(\sigma=0)$ is  real, one has 
\begin{align}
	\mathrm{Re}[\zeta(\sigma=1)-\zeta(\sigma=0)]=\frac{1}{2} \ln \left| \frac{\Omega_{+}}{\omega_{+}} \right |, \label{Eq64}
\end{align}
and
\begin{align}
	\mathrm{Im}[\zeta(\sigma=1)]=\frac{1}{2} \tan ^{-1} \frac{\mathrm{Im}[\Omega_{+}]}{\mathrm{Re}[\Omega_{+}]}\, ,
\end{align}
where Eq.~\eqref{Eq54} has been used. 
 Therefore,  
the cosmological complexity $\mathcal{C}_{FS}$ can be obtained through calculating 
\begin{eqnarray}
	\mathcal{C}_{FS}&=& \frac{1}{2}\sqrt{\left(\ln \left |\frac{\Omega_{+}}{\omega_{+}}\right|\right)^{2} + \left( \tan^{-1}\frac{\text{Im} |\Omega_{+}|}{\text{Re} |\Omega_{+}|}\right)^2} \,.
\end{eqnarray}

\end{document}